\documentstyle[aas2pp4]{article}
\input psfig

\def\oviq{OVI?  $\lambda 1033$}
\def\lya{Ly$\alpha$ $\lambda 1216$}

\def\nv{NV $\lambda 1240$}
\def\nvq{NV? $\lambda 1240$}
\def\siiv{SiIV $\lambda 1397$}

\def\civ{CIV $\lambda 1549$}

\def\ciii{CIII] $\lambda 1909$}
\def\ciiiq{CIII]? $\lambda 1909$}
\def\ciiib{CIII] $\lambda 2326$}

\def\neivq{[NeIV]? $\lambda 2424$}
\def\mgii{MgII $\lambda 2798$}

\def\nev{[NeV] $\lambda 3426$}
\def\nevq{[NeV]? $\lambda 3426$}
\def\oii{[OII] $\lambda 3727$}
\def\oiiq{[OII]? $\lambda 3727$}

\def\neiiiq{[NeIII]? $\lambda 3868$}
\def\caiik{CaIIK $\lambda 3933$}
\def\caiih{CaIIH $\lambda 3968$}

\def\hd{H$\delta$ $\lambda 4101$}

\def\hg{H$\gamma$ $\lambda 4340$}

\def\hbq{H$\beta$? $\lambda 4861$}
\def\oiiia{[OIII] $\lambda 4959$} 
 
\def\oiiib{[OIII] $\lambda 5007$}
  
\def\ha{H$\alpha$ $\lambda 6563$}
\def\haq{H$\alpha$? $\lambda 6563$}
\def\nii{[NII] $\lambda 6583$}
\def\sii{[SII] $\lambda 6717$}

\twocolumn

\begin{document}

\title{Redshifts of CLASS Radio Sources}

\author{D.R. Marlow\altaffilmark{1}, D. Rusin}
\affil{ Department of Physics and Astronomy, University of Pennsylvania, 
209 S. 33rd St., Philadelphia, PA, 19104-6396}
\altaffiltext{1}{and NRAL, Jodrell Bank, University of Manchester, 
Macclesfield, Cheshire SK11 9DL, UK}

\author{N. Jackson, P.N. Wilkinson, I.W.A. Browne}
\affil{NRAL, Jodrell Bank, University of Manchester, Macclesfield, Cheshire
SK11 9DL, UK} 

\author{L. Koopmans}
\affil{Kapteyn Astronomical Institute, Postbus 800, NL-9700 AV Groningen, The
Netherlands} 

\begin{abstract}
Spectroscopic observations of a sample of 42 flat-spectrum radio sources from
the Cosmic Lens All-Sky Survey (CLASS) have yielded a mean redshift of $<z> =
1.27$ with an RMS spread of $0.95$, at a completeness level of $64\%$. The
sample consists of sources with a 5-GHz flux density of $25-50$ mJy, making it
the faintest flat-spectrum radio sample for which the redshift distribution
has been studied. The spectra, obtained with the Willam Herschel Telescope
(WHT), consist mainly of broad-line quasars at $z>1$ and narrow-line galaxies
at $z<0.5$. Though the mean redshift of flat-spectrum radio sources exhibits
little variation over more than two orders of magnitude in radio flux density,
there is evidence for a decreasing fraction of quasars at weaker flux
levels. In this paper we present the results of our spectroscopic
observations, and discuss the implications for constraining cosmological
parameters with statistical analyses of the CLASS survey. 
    
\end{abstract}


\section{Introduction} \label{sec:intro}

The number of gravitational lens systems found in systematic surveys provides
a strong constraint on cosmological models, especially those with a
cosmological constant (Kochanek 1996a; Falco, Kochanek and Mu\~noz 1998,
hereafter FKM; Helbig et al.\ 1999). Statistical analyses of lens surveys
currently give a 2$\sigma$ upper limit on the cosmological constant in flat
models ($\Omega_{\Lambda}$+$\Omega_{M}$=1) of $\Omega_{\Lambda} < 0.62$
(FKM). These constraints are consistent with those obtained from Type Ia
supernovae (eg.\ Perlmutter et al.\ 1999).

Conducting a lens survey at radio frequencies has several advantages. Most
importantly, radio lens searches do not suffer from the observational biases
that plague optical surveys, particularly those due to dust extinction in the
lensing galaxy. In addition, a survey of compact radio sources with the Very
Large Array (VLA) provides maps with consistently high resolution
($0\farcs25$) and dynamic range ($\sim 100$), allowing lens systems with even
large component flux density ratios ($\sim 10:1$) to be easily identified in
most snapshot observations. The short integration time required for such
sources ($30$ seconds) means that large samples can be mapped in any given
observing run.

The drawback of radio lens surveys is that they are conducted in flux
density ranges where there is little information on the source redshift
distribution.  Complete redshift surveys of radio sources exist only for those
with $S_{\rm{5~GHz}}>$ 300 mJy (eg.\ the CJ samples, Henstock et al.\ 1997; and
the Parkes Half-Jansky Flat-Spectrum Survey (PHFS), Drinkwater et al.\ 1997). 
The Cosmic Lens All-Sky Survey (CLASS; Myers et al.\ 2000), the largest and
most successful lens survey to date, has a flux density limit of
$S_{\rm{5~GHz}}\geq30$ mJy. Kochanek (1996b) showed that a lack of redshift
information at these flux levels will lead to serious systematic uncertainties
in the derived cosmological constraints. FKM attempted to address this problem
with three samples of flat-spectrum radio sources selected from the
MIT-Greenbank Survey (Stern et al.\ 1999) and the Jodrell-VLA Astrometric
Survey (JVAS; Patnaik et al.\ 1992; Browne et al.\ 1998; Wilkinson et al.\
1998; King et al.\ 1999). The 5-GHz flux density ranges of these samples were
$50-100$, $100-200$ and $200-250$ mJy, respectively. They found a mean source
redshift of $<z>\sim1.2$, with a decreasing fraction of identified quasars at
weaker flux levels, falling to less than $50\%$ of sources at 100 mJy. 

To extend the spectroscopic study of flat-spectrum radio sources to even
weaker flux densities, a program of redshift determination for a small sample
of CLASS sources was undertaken. The aim was to obtain information on the
redshift distribution of the unlensed sources in CLASS. The same source
population comprises the ``parent'' sample of lensed sources in the JVAS
survey. This paper describes the selection and observation of the CLASS
optical sample and discusses some preliminary results.  

\section{Sample Selection}

The selection of the CLASS spectroscopic sample was made from the GB6 5-GHz
radio survey (Gregory et al.\ 1996). Those sources within the flux density
range  25 mJy $\leq S_{\rm{5~GHz}} \leq$ 50 mJy in the sky regions $17^{h} \leq
\alpha \leq 18^{h}$, $55^{\circ} \leq \delta \leq 70^{\circ}$ and $01^{h} \leq
\alpha \leq 02^{h}$, $35^{\circ} \leq \delta \leq 45^{\circ}$ (B1950.0) 
were selected. To restrict the sample solely to flat-spectrum sources, the
Westerbork Northern Sky Survey (WENSS; Rengelink et al.\ 1997) at 327 MHz was
used to determine two-point spectral indices for the GB6-selected sample. Due
to the size of the GB6 beam ($3.7'\times3.3'$) compared to that of WENSS 
($54''\times54'' \rm{cosec} \ \delta$), the correlation was limited to
those sources within two arcminutes of the GB6 pointing position. 
Sources with two-point spectral indices between 327 MHz and 5 GHz 
of $\alpha\geq -0.5$ (where $S\propto\nu^{\alpha}$) were selected. In
addition, those sources detected in the GB6 survey but not in the WENSS survey
were included in the sample, since they are likely to have inverted spectra
that decrease rapidly towards lower frequencies, and therefore fall below the
WENSS 327-MHz limiting flux density of 18 mJy. 

Only those sources that had previously been observed and detected in the
CLASS VLA observations of 1994 and 1995 were selected for the final sample. By
including this restriction in the sample definition, accurate VLA radio
positions could be used to identify the sources on the Palomar Sky Survey
(POSS). The result of this selection process was a sample of 42 radio sources
from CLASS.

\section{Observations}

Observations were carried out with the William Herschel Telescope (WHT) at La
Palma and the ISIS Spectrograph. The ISIS detector uses a $1124 \times 1124$
pixel CCD, with a 24$\ \mu$m pixel size. The ISIS R158 red and blue gratings
were used, dispersing light into red and blue orders covering the wavelength
range from 3000--10000\AA. The observations took place over four nights from
1997 August 2--5. A total of 40 CLASS objects were observed. The remaining two
objects (0157+41 and 1715+57) were identified with bright galaxies that had
previously determined redshifts in the literature.

Each source was observed for 200--600 seconds over the first two nights. The
fainter sources for which no emission lines were readily detected were
followed up during the next two nights with integrations of up to 5400 seconds.
The slit width used varied from $1\farcs0$ to $2\farcs0$ depending on the
observing conditions. Overall the weather was very good with little dust or
cloud cover, and an average seeing of $\sim 0\farcs8$. The majority of the
targets were visible on the finder-TV, and were identified using bright nearby
objects on finding charts generated from POSS. However, approximately
one-third of the sources were not visible and had to be located by ``blind
offsetting'', which resulted in a detection rate of $50\%$. Misidentifications
due to crowding are negligible.

\section{Data Reduction and Analysis}

The data were analyzed with the IRAF\footnote{ IRAF is distributed by the
National Optical Astronomy Observatories, which are operated by the
Association of Universities for Research in Astronomy Inc. under cooperative
agreement with the National Science Foundation.} data reduction package. First
the bias level was removed from the data frames using the overscan region of
the chip and bias frames. Flat-fielding was then performed. The curvature of
the two-dimensional data frames was traced, and the one-dimensional spectra
were extracted.
 
The extracted spectra were wavelength-calibrated using a single Cu-Ne/Cu-Ar
arc for each of the two wavelength orders. The arcs were selected from the end
of each observing run, to minimize the effects of shifts in the wavelength
scale. The spectra were flux-calibrated using observations of
spectrophotometric standard stars (Oke and Gunn 1983). This was done by
determining a sensitivity function relating counts to flux density as a
function of wavelength, which was then applied to the spectra. An observation
of BLLac was performed and the spectra were normalized by the featureless
spectrum.

The flux-calibrated spectra were trimmed to remove the wavelength
extremes where the sensitivity is low and hence where the spectra
are noisy and badly calibrated. The two orders were then combined to
give spectra covering the wavelength ranges 3300--6100\AA\ and
6300--8300\AA. 


The majority of the sources show one or more emission or absorption lines. The
remaining spectra which show no obvious lines can be divided into two
categories: those that have insufficient signal-to-noise to show any reliable
features, and those that have relatively good signal-to-noise but show no
distinguishing lines and hence have BLLac-type spectra.

For the spectra with more than one identified line, the source redshift was
determined from an unweighted mean of the redshifts of the individual
lines. The error on the redshift was calculated from a combination of the mean
errors from the Gaussian fitting procedure on the emission and absorption
lines. In the few cases where only one line was identified, the redshift is
simply the line redshift, and its error simply the error derived from the
line fitting. In addition, a systematic uncertainty due to the
wavelength calibration should be folded into the error on the redshift. This
is of the order 5--10\AA\ for the different spectra, which is roughly
the same level of error associated with the fitting. This source of
uncertainty has not been included in the quoted redshift errors.

\section{Results}

Table 1 lists the name, radio position, GB6 5-GHz flux density, total exposure
time, redshift, spectral type (galaxy or quasar), and identified emission and
absorption lines for each of the 42 sources in the CLASS sample. Plots of the
individual spectra are available at the CERES homepage.\footnote{ 
http://www.jb.man.ac.uk/$\sim$ceres1} 

Assuming all tentative redshifts to be correct, the CLASS sample has a mean
redshift of $<z>=1.27$ and an RMS spread of 0.95, at a completeness level of
$64\%$. Fig. 1 shows the mean redshifts for a variety of samples of
flat-spectrum radio sources, spanning more than two orders of magnitude in
flux density. There appears to be little evidence for a decrease or increase
in the mean redshift of such sources with decreasing radio flux. 

\begin{figure}[ht!]
\hspace{1 cm}\psfig{file=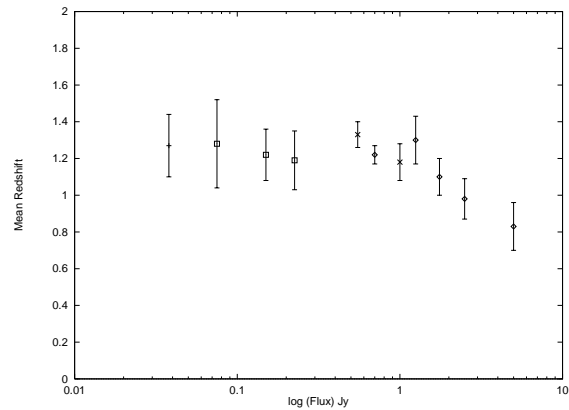,width=3in,angle=270} 
\caption{Mean source redshift as a function of radio flux density.
The observed mean values are plotted for the CLASS (cross), FKM (squares), CJ
(Xs) and PHFS (diamonds) samples. The error bars indicate the $1\sigma$ error
on the mean redshift.} 
\end{figure}

\begin{figure}[ht!]
\hspace{1 cm}\psfig{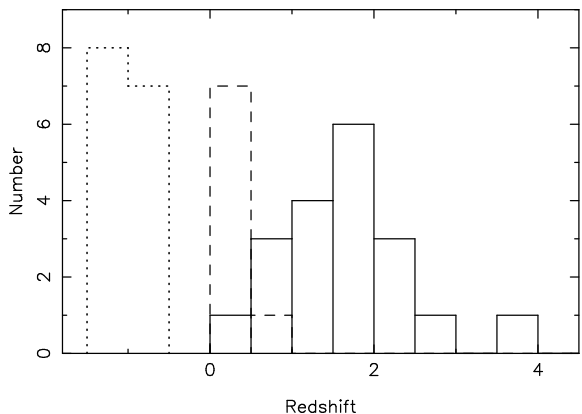} 
\caption{Redshifts of the CLASS sample. The redshifts from galaxies are given
in dashed lines. Those from quasars are given in solid lines. The values at
negative redshifts refer to objects with undetermined redshifts.}
\end{figure}

The CLASS optical sample includes 19 quasars, 8 galaxies and 15 unidentified
sources. Four of the unidentified sources were clearly detected and two were
weakly detected in the red only. Fig. 2 shows a histogram of the redshifts
obtained for the CLASS sample. The sources are strongly segregated into quasars
with broad emission lines at $z>1$ and narrow-line galaxies at $z<0.5$.

The relative fractions of quasars and galaxies in various flat-spectrum
samples are plotted in Fig. 3. In the brightest samples, approximately $80\%$
of the sources are quasars. At the lower flux densities of the CLASS sample,
quasars comprise less than $50\%$ of the sources. The fraction of identified
low-redshift galaxies remains quite constant at $\sim 20\%$ from $30$ mJy to
$1$ Jy. This may indicate that the unidentified sources at low flux densities
are drawn from a different population than the identified galaxies, and are
perhaps at a higher redshift. If this is the case, the six unidentified
sources that were detected may have lines beyond the long wavelength cut-off
of our spectra, while the remaining sources are just too faint to be detected
at their redshifts.

\begin{figure}[ht!]
\hspace{1 cm}\psfig{file=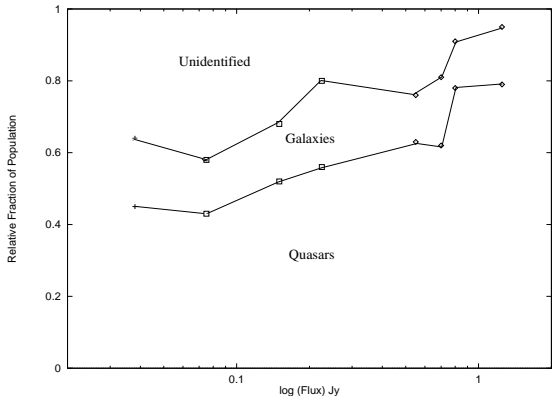,width=3in,angle=270} 
\caption{Relative source populations for 
flat-spectrum radio samples. The distribution of galaxies and quasars 
as a function of radio flux density is shown for the CLASS, FKM and PHFS
samples.} 
\end{figure}

Sources that are lensed into the CLASS survey come from an even fainter parent
sample of flat-spectrum sources, typically in the 5--30 mJy range. If the
above trends continue down into this regime, with fewer quasars and more
galaxies, one would expect to see lensed extended emission from the
host galaxies in CLASS lens systems. This is consistent with the detection of
arcs in the optical and infrared images of several JVAS and CLASS lenses
(MGJ0414+0534, Falco et al.\ 1997; B1359+154, Rusin et al.\ 2000; B1933+507,
Marlow et al.\ 1999; B1938+666, King et al.\ 1998). Each of these lensed
objects is at $z>2$, which may bolster the argument that the unidentified
sources in the CLASS optical sample are galaxies at high redshift.

\section{Discussion}

The number of gravitational lens systems found in a well-defined sample of
sources can place strong limits on the cosmological constant. This is done by
comparing the observed lensing rate with the predictions of various
cosmologies. The optical depth for lensing depends both on the cosmological
parameters and the redshifts of the sources being studied. Systematic
uncertainties in the statistical analyses of radio lens surveys are therefore
created by the paucity of data on the redshift distribution of unlensed
sources fainter than 300 mJy.

Recently Helbig et al.\ (1999) analyzed the lensing rate in the JVAS
statistical sample, which includes four lenses and approximately 2300 radio
sources. At $95\%$ confidence, the lower and upper limits on
$\Omega_{\Lambda}$ in a flat universe are $-0.85$ and $0.84$,
respectively. These results are consistent with other recent measurements of
$\Omega_{\Lambda}$, both from lensing statistics and alternate cosmological
probes such as Type Ia supernovae (Perlmutter et al.\ 1999). Helbig et
al.\ assumed that the redshift distribution of the JVAS survey was identical to
that of the CJ surveys, which were done at a higher flux density level. A
similar assumption is harder to justify for CLASS, which searches for lensing
among sources that are fainter than the CJ sample by more than an order of
magnitude.

Future statistical analyses of CLASS require a determination of the redshift
distribution of flat-spectrum radio sources down to the $30$ mJy level. Our
current results only partially address this issue. The measured redshifts
of the CLASS sources appear to be consistent with those of brighter radio
samples ($<z> \approx 1.2 \pm 0.1$). However, the mean redshift may change
once completeness is improved, and the nature of the currently unidentified
sources is determined. 

An underestimate of the mean redshift of the unlensed source population would
lead to a overestimate of the cosmological constant, and vice versa.
Consequently, the uncertainty in the redshift distribution of the
CLASS sample permits large variations in the expected number of lenses for a
given cosmological model. This is nicely illustrated by Kochanek (1996b).  For
radio luminosity functions constrained to produce four JVAS lenses, the mean
redshift of a 50 mJy source varies from 0.4 for $\Omega_{M}=0$, to 1.9 for
$\Omega_{M}=1$, to almost 4.0 for $\Omega_{M}=2$ in a flat universe. Although
these are extreme values, they underscore the importance of understanding the
source redshift distribution when using the observed lensing rate as a probe
of the cosmology.

At this time, the constraints on cosmological parameters from the radio and
optical lens surveys (and other tests) appear consistent, suggesting that the
observational uncertainties are not overwhelming. The CLASS survey, however,
pushes down into a lower flux density regime in which very little is known
about the source redshift distribution. Therefore, a reliable and complete
spectroscopic study of flat-spectrum radio sources for the entire flux density
range from 30--300 mJy must be obtained if CLASS is to fulfill its promise as
a cosmological tool. To this end, a new program of spectroscopic observations
is currently being undertaken by the CLASS/CERES collaboration at La Palma.

\acknowledgements 

We thank Emily Xanthopoulos for making the spectra available online.  This
research is based on observations made with William Herschel Telescope
operated on the island of La Palma by the Isaac Newton Group in the Spanish
Observatorio del Roque de los Muchachos of the Instituto de Astrofisica de
Canarias. This research also made use of the NASA/IPAC Extragalactic Database
(NED) which is operated by the Jet Propulsion Laboratory, Caltech, under
contract with the National Aeronautics and Space Administration.  The National
Radio Astronomy Laboratory is a facility of the National Science Foundation
operated under cooperative agreement by Associated Universities Inc. This
research was supported in part by European Commission TMR Programme, Research
Network Contract ERBFMRXCT96-0034 ``CERES''.

\begin{table*}
\begin{tabular}{ ccccccccc }
\hline\hline
Source  &  R.A. & Decl. & Flux& Integ.  & Redshift &Q/G & {Line ID} 
\\
  & (J2000) & (J2000) & (mJy) & (s) & $z$ &  \\
(1)&(2)&(3)&(4)&(5)&(6)&(7)&(8)\\
\hline
 & & & & & &  &  \\
 0103+43  & 01 03 28.08 & 43 22 59.5 & 41  &  1800  &  -- & -- &
Detected/  \\ 
 & & & & & &  & Featureless \\
 & & & & & &  &  \\
 0105+39  & 01 05 09.30 & 39 28 15.3& 48&  3000  & (0.083$\pm0.001$)   &G?
& \oiiq   \\ 
 & & & & & & & \haq  \\
 & & & & & &  &  \\
 0106+44  & 01 06 21.40 & 44 22 27.4 & 47  &  3600  & -- & -- &
Featureless    
 \\ 
 & & & & & &  &  \\ 
 0123+3806  & 01 23 28.87& 38 06 36.7&40 &300&1.656$\pm0.004$&Q & \civ  \\ 
           & & &     &        &  &    & \ciii     \\ 
 & & & & & & &  \mgii \\
& & & & & &  &  \\ 
 0123+3843  & 01 23 23.83 & 38 43 57.9 & 27  &  4800  &-- & -- &
Detected in red/
   \\ 
 & & & & & &  & Featureless\\ 
 & & & & & &  &  \\

0128+40  & 01 28 13.64 & 40 03 29.9 &47 &  1800  & 3.525$\pm0.003$& Q  & \oviq \\
& & & & & &  &  \lya\\  
            & & &     &        &                   &      &      \nv  \\
           & & &     &        &                    &     &      \civ\\ 
& & & & & &  &  \\ 
0128+44  & 01 28 41.34 & 44 39 17.9 & 27 &2700& 0.228$\pm0.001$ & G &\oii
\\ 
           & & &     &        &                        &     &  \ha    \\
          & & &     &        &                        &     &  \sii   \\
& & & & & &  &  \\ 
0131+44  & 01 31 03.36 & 44 28 01.9 &27 &1200& 1.123$\pm0.001$&Q & \ciii
\\ 
& & & & & &  & \neivq \\ 
          & & &     &        &                        &    &   \mgii  \\   
& & & & & &  &  \\
 0132+43  & 01 32 06.06& 43 45 34.4 & 41&  1200  &  1.812$\pm0.001$& Q  &\lya\\
            & & &     &        &                        &    &    \civ \\
           & & &     &        &                        &     &  \ciii  \\
           & & &     &        &                        &     &  \mgii \\ 

& & & & & &  &  \\ 
 0136+35  & 01 36 43.68 & 35 45 31.2 & 33 &  1800  &  1.871$\pm0.001$& Q
&  \lya\\ 
& & & & & &  & \nvq \\
            & & &     &        &                        &  &      \siiv \\
           & & &     &        &                        &   &    \civ \\
           & & &     &        &                        &   &    \ciii \\ 
& & & & & &  &  \ciiib \\
           & & &     &        &                        &   &    \mgii  \\  
& & & & & &  &  \\
 
\hline
\end{tabular}
\caption[ Table of CLASS redshifts.]{
Table of CLASS redshifts. Column (1) -- Source name.
Column (2) -- Right Ascension (J2000).
Column (3) -- Declination (J2000).
Column (4) -- Total flux density in the GB6 5-GHz catalogue.
Column (5) -- CCD exposure time in seconds.
Column (6) -- Redshift with error from Gaussian fitting. (Tentative
redshifts are given in brackets.)
Column (7) -- Galaxy or Quasar-type spectrum.
Column (8) -- Identified lines.}
\normalsize
\end{table*}
\newpage
\setcounter{table}{0}
\begin{table*}
\begin{tabular}{ cccccccccc }
\hline\hline
Source  &  R.A. & Decl. & Flux& Integ.  & Redshift &G/Q & {Line ID} 
\\
  & (J2000) & (J2000) & (mJy) & (s) & $z$ &  \\
(1)&(2)&(3)&(4)&(5)&(6)&(7)&(8)\\
\hline
 & & & & & &   \\
 0136+44  & 01 36 47.32& 44 01 10.2 &36&  1200  & --& -- &  Featureless  
    \\ 
 & & & & & &   & \\ 
 0137+39  & 01 37 45.14 & 39 41 36.1 & 38&3000 & (1.396$\pm0.004$)&Q? &
\ciiiq  
\\ 
& & & & & & &  \nevq\\
& & & & & & & \\  
 0143+37  &  01 43 02.54 & 37 05 16.1 & 46 &  1000  &  --    & -- &
Featureless
      \\
 & & & & & &  &  \\
 0151+43  & 01 51 18.38 & 43 32 00.5 &34  &  1200  &  2.192$\pm0.002$ &Q &      
\lya   \\
 & & & & & &  & \nv\\ 
             & & &     &        &                        &   &     \siiv
\\
            & & &     &        &                        &    &    \civ  \\
 & & & & & &  & \\ 
 0151+44  & 01 51 20.88 & 44 17 35.9 & 44 &  600  & 1.976$\pm0.003$     &
Q &   
\nv    \\ 
             & & &     &        &                        &     &   \civ
\\ 
& & & & & &  & \\
 0154+35  & 01 54 45.46 & 35 58 04.6 & 34 &   1200  &--  & --  &
Detected in red/    
  \\ 
 & & & & & &  & Featureless \\
 & & & & & &  &  \\
 0154+44  &  01 54 54.47 & 44 33 37.9 & 40  &  3300  &  --   &--  &
Detected/
 \\ 
 & & & & & &  & Featureless\\
 & & & & & &  &  \\
 0156+44  & 01 56 28.52 & 44 59 56.5 & 46 &  600  & (0.214$\pm0.001)$& G?&
\oiiq
 \\ 
 & & & & & &   & \\
0157+41  & 01 57 05.01 & 41 20 30.6 & 30 & -- & 0.0811 &G& Abell
0276 \\
 & & & & & & & \\
 0159+41  & 01 59 49.33 & 41 44 32.9 & 33  &  1200  &   --   &-- &
Featureless  
   \\
 & & & & & & & \\
 1702+55  & 17 02 34.56 & 55 11 12.4 &34 &  600  & --  &--& Featureless
\\ 
 & & & & & && \\ 
 1702+58  & 17 02 41.37 & 58 13 10.1 & 38&  5400  & --  & --&Featureless
\\ 
 & & & & & & &\\ 
 1703+63  & 17 03 04.72 & 63 42 29.8 & 34&  3600  & --    &--
&Featureless\\
 & & & & & & &\\
 1706+60  & 17 06 56.08 & 60 23 37.9 & 27&  600  & 1.472$\pm0.002$& Q
& \civ\\ 
 & & & & & & &\ciii\\ 
& & & & & & &\\
 1715+57  & 17 15 22.98 & 57 24 40.3 & 35  & --   & 0.0279&G
&NGC6338  \\
 & & & & & & \\
 1715+63  & 17 15 35.96 & 63 23 36.0 & 37&  600  &  2.185$\pm0.001$  & Q &
\lya\\ 
& & & & & &  &\siiv \\
& & & & & &   &\civ \\ 
& & & & & &  & \ciiiq \\
& & & & & & & \\
 1721+59  & 17 21 00.65 & 59 26 49.4 & 41 &  1200  & 0.587$\pm0.001$   & G
&   \oii\\
& & & & & &  & \oiiia\\
& & & & & &  &\oiiib \\
& & & & & & & \\
 \hline
\end{tabular}
\normalsize
\caption[Table of CLASS redshifts (continued).]{Table of CLASS redshifts (continued).}
\end{table*}
\setcounter{table}{0}
\newpage
\begin{table*}
\begin{tabular}{ cccccccccc }
\hline\hline
Source  &  R.A. & Decl. & Flux& Integ.  & Redshift &G/Q& {Line ID} 
\\
  & (J2000) & (J2000) & (mJy) & (s) & $z$ &  \\
(1)&(2)&(3)&(4)&(5)&(6)&(7)&(8)\\
\hline
 & & & & & &  \\
 1728+55  & 17 28 11.64 & 55 32 30.5 & 42 &  600  & 1.404$\pm0.002$   & Q
&    
\civ\\
 & & & & & &  & \ciii \\
& & & & & &   & \mgii \\
 & & & & & & & \\
 1728+67  & 17 28 21.67 & 67 53 34.2 & 27&  1200  &1.987$\pm0.004$ &Q  &
\lya  \\
& & & & & & &  \siiv\\  
 & & & & & & & \civ \\
 & & & & & & & \ciii \\
 & & & & & & &  \\
 1729+67  & 17 29 20.47 & 67 02 13.0 & 26 &  1800  & 0.952$\pm0.001$   & Q
& \ciii \\
& & & & & &  &\mgii \\
& & & & & &  & \nev \\
& & & & & &  & \oii \\ 
& & & & & &  & \neiiiq \\   
& & & & & &  &\\ 
 1730+60  & 17 30 52.71 & 60 25 16.7 &  36&   600  & 0.730$\pm0.001$   & Q
&  \mgii \\ 
& & & & & &  & \hd\\
& & & & & & &  \hg \\  
& & & & & &  & \\
 1732+55  & 17 32 23.74 & 55 24 52.8 & 32&  300  & 0.064$\pm0.001$    & G
&  \caiih \\
& & & & & &  & \caiik\\ 
& & & & & & &  \ha\\
& & & & & & & \\ 
 1733+58  & 17 33 07.53& 58 19 47.6 & 26 &  600  & 1.631$\pm0.002$   &  Q
&  \civ  \\
& & & & & &  & \ciii \\
& & & & & &  & \mgii \\
 & & & & & & &  \\
 1733+60  & 17 33 51.33 & 60 49 34.0 &26  &  1200  & -- & -- &
Featureless\\ 
 & & & & & &  & \\
 1735+56  & 17 35 13.78 & 56 50 21.8  &31 &  5400  & --  & -- &
Detected/   \\
 & & & & & &  & Featureless \\
 & & & & & &  & \\
 1745+68  & 17 45 33.18 & 68 56 05.7  &43 & 4800  & --  &--  &
Featureless  \\ 
 & & & & & &  & \\
 1747+59  & 17 47 33.94 & 59 02 47.9  & 25 & 1800  &0.981$\pm0.001$  & Q &
\ciii \\ 
& & & & & &  & \mgii\\ 
& & & & & & &\\
 1756+58  &  17 56 29.14 & 58 06 58.2 & 38 &   200  & 0.192$\pm0.002$
&G  &  \oii\\
& & & & & & &  \ha \\ 
& & & & & &  & \nii \\  
 & & & & & &  &\sii \\
& & & & & &  & \\ 
 1756+59  & 17 56 11.83 & 59 18 56.8 &25  &   3000  & (0.479$\pm0.001$) &
Q    & \mgii \\ 
& & & & & & &  \hbq \\ 
& & & & & & &  \\

\hline
\end{tabular}
\normalsize
\caption[Table of CLASS redshifts (continued).]
{Table of CLASS redshifts (continued).}
\end{table*}
\newpage
\setcounter{table}{0}

\begin{table*}
\begin{tabular}{ cccccccccc }
\hline\hline

Source  &  R.A. & Decl. & Flux& Integ.  & Redshift &G/Q& {Line ID} 
\\
  & (J2000) & (J2000) & (mJy) & (s) & $z$ &  \\
(1)&(2)&(3)&(4)&(5)&(6)&(7)&(8)\\
\hline
 & & & & & &  \\
 1756+65  & 17 56 40.36 & 65 31 45.9 &  43 &   1200  &2.790$\pm0.001$&  Q
&   \lya \\
& & & & & & & \nv\\
& & & & & &  & \civ \\
& & & & & &  & \ciii \\ 
 & & & & & &  &\\
 1759+64  & 17 59 23.01 & 64 48 56.6 &33  &   1200  & 2.476$\pm0.002$& Q
& \lya \\ 
& & & & & & & \nv\\
& & & & & &  & \siiv\\
& & & & & &  & \civ\\ 
& & & & & & &  \ciii \\ 
& & & & & & & \\
 1800+61  & 18 00 25.24 & 61 26 15.1 &30 &  3000  & -- & --& Detected/
\\
 & & & & & &  & Featureless \\ 
 & & & & & &  & \\
 & & & & & &  & \\
\hline
\end{tabular}
\normalsize
\caption[Table of CLASS redshifts (continued).]
{Table of CLASS redshifts (continued).}
\end{table*}
\normalsize

\end{document}